\documentstyle[floats,prd,aps]{revtex}

\input psfig.sty
\newcommand{\be}{\begin{equation}}
\newcommand{\ba}{\begin{eqnarray}}
\newcommand{\ee}{\end{equation}}
\newcommand{\ea}{\end{eqnarray}}

\begin{document}
\title{Fractals and Scars on a Compact Octagon}
\author{Janna Levin and John D. Barrow}
\address{DAMTP, Cambridge University,
Silver St., Cambridge CB3 9EW}
\address{and}
\address{Astronomy Centre, University of Sussex,
Brighton BN1 9QJ, U.K.}
\maketitle

\begin{abstract}

A finite universe naturally
supports chaotic classical motion.
An ordered fractal emerges from the chaotic dynamics which we
characterize in full for a compact 2-dimensional octagon.
In the classical to quantum transition, the underlying fractal can
persist in the form of scars, ridges of enhanced amplitude in the
semiclassical wave function.
Although the scarring is weak on the octagon, we suggest possible
subtle implications of fractals and scars in a finite universe.

\end{abstract}

\vskip 20pt
\indent{05.45.-a,05.45.Df,05.45.Mt,98.80.Cq}

\vskip 10truept

\widetext

\twocolumn
\narrowtext
\begin{picture}(0,0)
\put(410,190){{ }}
\end{picture} \vspace*{-0.15 in}

Chaos is often avoided in fundamental physics and cosmology.  
A recent surge of interest in a finite universe
has highlighted a well known chaotic system, a compact negatively
curved space \cite{cqg}.
Most efforts in cosmology have struggled to circumvent the inherent chaos.
Instead of avoiding chaos, we investigate the structure of the chaotic
flows on a compact space.
Since we are able to deduce analytically 
many of the chaotic properties, we concentrate
on the compact 2d octagon and
study the
classical motion and the emergent fractal properties.
Poincar\'e realized that the periodic orbits, though seemingly
special, define the entire structure of a dynamical system.
With the advent of chaos, periodic orbits grow
exponentially in number as a function of their length.  The
invariant set of periodic trajectories
packs itself into a fractal to
accommodate the dense proliferation of orbits.
We derive a very precise
counting of the periodic orbits, determine the 
topological entropy, derive a recurrence relation for all of the
fixed points, and analytically compute a spectrum of fractal dimensions. 
We then numerically confirm our analytic results by generating the fixed
points and measuring the dimensions of the set.

The fractals may not be effaced in 
the classical to quantum transition.
A phenomenon known as scarring of the quantum wave function
has been witnessed in 
systems which are classically chaotic \cite{heller1}.  The scars refer to
ridges along which the amplitude of the 
semiclassical wave function is enhanced.  
We argue that the scars can be interpreted as the smooth remnant of
the underlying classical fractal.  
We 
relate the closed orbits on the 2d space to the occurrence
of scars.  
The compact octagon is so thoroughly chaotically mixed that the scarring
is minimal, consistent with the fractal properties as discussed.
In a finite model of the universe, the faint scars could 
seed a
weblike distribution of matter threaded around the space.

Geodesic motion on a compact $2d$ pseudosphere is  
completely chaotic.  
On the Poincar\'e unit disc the metric can be written
	\be
	ds^2=-d\eta^2+{4(1-r^2)^{-2}}\left (dr^2+r^2d\phi^2\right )
	.
	\ee
Geodesics are semi-circles which
are orthogonal to the boundary at $r=1$.
We consider null geodesics for which 
the coordinates $(r,\phi)$ have the conjugate momenta
	\be
	(\Pi_r,
	\Pi_\phi)={4 (1-r^2 )^{-2}}(\dot r, r^2\dot \phi )
	.
	\ee
There are two conserved quantities. 
Since $\dot \Pi_\phi=0$,
there is a conserved angular momentum
	\be
	L={4 (1-r^2 )^{-2}}r^2\dot \phi
	\ee
and the kinetic energy is conserved
	\be
	{1\over 2}\left (\dot r^2+r^2\dot\phi^2\right )
	{4 (1-r^2 )^{-2}}
	={1\over 2}
	.
	\label{ke}
	\ee
Condition (\ref{ke}) is the same as
the restriction that photons travel at unit speed.

\begin{figure}[tbp]
\centerline{{
\psfig{file=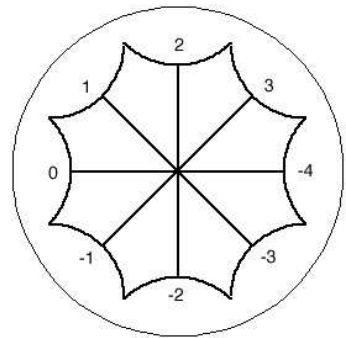,width=1.75in}}} \vskip 15truept
\caption{The symmetric octagon drawn in the Poincar\'e unit disc.  The
octagon forms a two-handled surface when the opposite sides are glued together.
Each edge corresponds to a generator $g_j$ with $j=-4,..,3$ and 
are numbered accordingly.}
\label{fund}
\end{figure}

We consider the simplest compact surface of negative curvature with genus 2.
The fundamental domain is a symmetric regular octagon with the
opposite edges identified in pairs
as in fig.\ \ref{fund}.
The identifications are effected by a discrete subgroup of the Lorentz
group.
The horizontal identifying boost can be represented 
in coordinates $x=r\cos \phi$ and $y=r\sin \phi$ as
	\be
	g_0=\pmatrix{\cosh \tau_1/2 & \sinh \tau_1/2 \cr
	\sinh \tau_1/2 & \cosh \tau_1/2  } 
	\label{go}
	\ee
with $	\cosh\tau_1/2=1+\sqrt{2}$,
while the other four are related by a rotation of 
$g_0$ through the angle $j\pi/4 $.
The four generators are related to their respective
inverses by a rotation through $\pi$.
We denote all the generators and their inverses compactly as
$g_j$ with $j=-4,..,3$ so that the $g_0g_{-4}=g_{3}g_{-1}=
g_{2}g_{-2}=g_{1}g_{-3}={\bf I}$.  These generators obey the one relation
	\be
	g_0g_{-3}g_2g_{-1}g_{-4}g_1g_{-2}g_3={\bf I}
	.\label{rela}
	\ee

\begin{figure}[tbp]
\centerline{{
\psfig{file=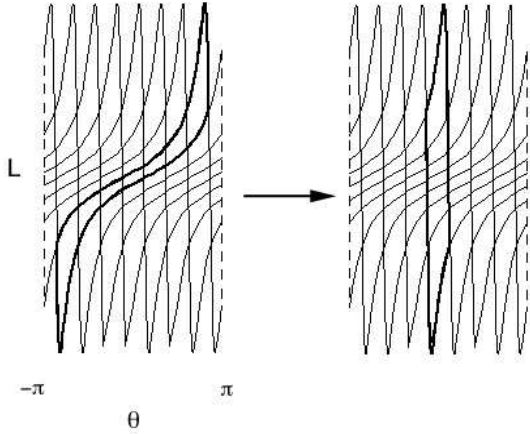,width=3.in}}}
\caption{The action of the map on the exit-entry domains for
geodesics.  
An orbit is completely
specified by the angular momentum $L$ and the angular coordinate
$\phi(t=0)=\theta $ at which the geodesics intersects the boundary in
the past.  All the orbits within the darkened exit domain above
exit edge 0 of fig.\ \ref{fund}.
The map takes this exit domain 
into the darkened vertical entry domain.
}
\label{map}
\end{figure}

A geodesic is uniquely specified by two parameters $(\theta,L)$ where
$\theta $ is the angular coordinate of the geodesic as it intersects
the boundary in the past.
Balasz and Voros  \cite{bv}
developed a reentry map which takes initial values for 
$(\theta,L)$ through the appropriate edge and back into the
fundamental domain.   
The map which iterates all values through edge 0 of
fig.\ \ref{fund}
is
	\ba
	\tan(\theta_{n+1}/2) &=& e^{-\tau_1}\tan(\theta_n/2)	\label{mapeq}
	\\
	L_{n+1}= (\cosh\tau_1&+&\cos\theta_n
	\sinh\tau_1)L_n-\sinh\tau_1\sin\theta_n .\nonumber 
	\ea
The reentry map for orbits exiting any of the $8$ edges is effected by 
shifting $\theta \rightarrow \theta +j\pi/4 $ for $j=-4,...3$.
The map is Hamiltonian and 
area preserving.

A trajectory is a point in the plane $(\theta,L)$ and those 
trajectories lying
within the intersection of the two darkened separating curves of fig.\
\ref{map} exit 
edge 0.  
A separating curve is
	\be L_0={\sin(\theta-\pi/8)\over
	\cos(\theta-\pi/8)-\coth\tau_2} \label{sep}\ee 
with
$\coth\tau_2=2^{-5/4}(1+\sqrt{2})$.  
The other $7$ exit domains are shifted by $\pi/4$ as
represented in fig.\ \ref{map}.
All eight curves, $L_j$, are obtained from eqn.\ (\ref{sep}) by
shifting 
$\theta\rightarrow \theta +j\pi/4$ with $j=-4,..0,..3$.  The
map acts very much like an eightfold Baker's map.  The 
$8$ sections exit their respective edges and reenter as the equal-area,
nearly-vertical strips.  Each of the $8$ exit strips overlaps with
only $7$ of the reentry strips.  In other words,
the points can reenter any of the
other edges except the one just exited.
The original area is 
chopped into $8\times 7$ smaller areas, each of
which continues to flow through the space and gets resegmented into 
$8\times 7\times 7$ new areas.  The total number of pieces at order
$n$ is then $8\times 7^{n-1}$.
Covering the area with boxes of size
$\epsilon=(1/7)^{n/2}$ gives the box-counting dimension
$D_0=\lim_{n\rightarrow \infty}
\ln(7^{n-1})/\ln(7^{n/2})= 2$.  The fractal fills the allowed 
area.

The number of periodic orbits is slightly smaller than the number of 
areas.
The $g_j$ construct a symbolic alphabet out of which to build 
$n$-letter words.  
Each repeated word corresponds to a
periodic orbit.
The number of words\footnote{The number of words
actually counts fixed points and not periodic orbits.
For instance, the words  $g_1g_2$ and $g_2g_1$ describe two different 
fixed points along the same periodic orbit.  These both contribute to $N(n)$.
The number of periodic orbits is slightly less than this, $\sim N(n)/n$.}
 of order $n$, $N(n)$, is the sum of the number of words
which begin and end on the same letter, $O(n)$, and the number of
words which begin and end on different letters, $\Delta(n)$,
	\be
	N(n)=O(n)+\Delta(n)
	.
	\ee
A word cannot begin and end on inverse letters:
a repeated 
$n$-letter word of the
form
${g_ig_j...g_kg^{-1}_i}$ is equivalent to the repeated $(n-1)$-letter word
${g_j...g_k}$.
An $n$-letter word can be built from every $O(n-1)$ by attaching any
of the letters in the alphabet except the one which is inverse to 
the end letter.  So, there are $7 O(n-1)$ of these.
More $n$-letter words can be built from every $\Delta(n-1)$ by
attaching any of the letters in the alphabet except for the two
letters which are either the inverse of the end letter or the inverse of the
beginning letter.  So, there are $6 \Delta(n-1)$ of these.
Although $(n-1)$-letter words of the form $g_i....g^{-1}_i$ 
do not contribute to $N(n-1)$,
they do contribute to $N(n)$  (for $n\ge 3$)
by adding any letter
other than $g_i $ or $g^{-1}_i$.  The number of words
of order $(n-1)$ that 
begin and end on
inverse letters is 	
	\be 	I(n-1)=\sum_{m=2}^{m=n-2}\Delta(m) .\ee
It follows that 
	\be	N(n)=7O(n-1)+6\Delta(n-1)+6I(n-1)
	.\label{en}
	\ee
Eqn.\ (\ref{en}) can be simplified by noting 
that 
$O(n)=N(n-1)$.  We can combine these relations to reexpress (\ref{en}) for
$n\ge 3$ as
	\be
	N(n)=7N(n-2)+6N(n-1)-6N(1)
	.
	\label{recur}
	\ee
There is additional pruning at higher orders due to the one relation
eqn.\ (\ref{rela}).
However, this
pruning is extremely sparse and affects the counting very little.
The number in (\ref{recur}) 
is actually slightly smaller than $N(n)=7^n$ but tends
toward this behaviour as $n\rightarrow \infty$.
To see this, let $N\sim b^n$ in the limit of large $n$ and
solve for $b$ to obtain $N\rightarrow 7^n$.
The topological entropy is then, $H_T=\lim_{n\rightarrow
\infty} (1/n)\ln N=\ln 7$.

There is a spectrum of weighted dimensions which measures 
the distribution of frequencies and the distribution of the
points in space.
The spectrum is defined by covering the set with boxes of
size $\epsilon $ on a side and taking the limit as
as $\epsilon \rightarrow 0$,  
	\be
	D_q={1\over q-1}\lim_{\epsilon\rightarrow 0}
	{\ln\sum_{i=1}^{N(\epsilon)}p_i^q
	\over \ln \epsilon}.
	\label{specd}
	\ee
with $p_i$ the weight assigned to the $i$th box.  
If all closed orbits are equally
weighted then $p_i=1/N(\epsilon)$ and the fractal is self-affine
with the $D_q$ equal.  	
For a multifractal, some of the orbits are more popular than others and at
least some of the $D_q$ are unequal.

Following the reasoning of ref. \cite{ott} for the
Baker's map, we can use the areas of fig. \ref{map}
to analytically determine the spectrum of
dimensions for the map.
A given exit section of area $A$
overlaps with the reentry strip into $7$ overlap sections of area
$A_j$.
The map then takes these $7$ areas and lays them
down in nearly vertical strips preserving the area of each.
The natural measure on the vertical strip is then the relative area
$f_j=A_j/A$. The vertical strips will have nearly the same height and
differ only in relative width, so the relative width is also given by $f_j$.
There is a
simple scaling symmetry: if any of the vertical strips were to be
multiplied by the width of that strip,  
the features of the entire set are reproduced.  This kind of reproduction
on smaller and smaller scales is precisely what makes the
set 
fractal.  We can exploit the scaling to deduce the fractal dimension
\cite{ott}. 
In this vertical approximation the dimension is fractal only in the 
$\theta $ direction and $D_q=1+D_q^\theta$.
Write
the weighted sum of (\ref{specd}), $	I=\sum^{N(\epsilon)}p_i^q$,
as $I(\epsilon)=\sum_{j=1}^7 I_j$
with 
natural measure $f_j$ on the $j$th strip.
As we make $\epsilon $ smaller by $\epsilon/f_j$, we reproduce the 
structure of the entire set.  The scaling can be expressed as
	\be
	I_j(\epsilon)= f_j^q I(\epsilon/f_j)
 	.
	\ee
From the definition (\ref{specd}),
$I(\epsilon)\sim \epsilon^{(q-1)D^\theta_q}$, we obtain 
	\be
	\sum_{j=1}^7 f_j^{q+(1-q)D^\theta_q}=1
	.
	\label{app}
	\ee
Since the natural measures are normalized so that $\sum_jf_j=1$ it
must
be that the exponent $q+(1-q)D^\theta_q=1$ and it follows that
	\be
	D_q=1+D^\theta_q=2
	\label{dq}
	\ee 
for all $q$.  The fractal is self-similar and fills an area.
It should be noted that this dimension is computed by taking the
entire area filled with periodic trajectories of all lengths and
kneeding those areas into an iterate of parallel sections,
analogous to a Hamiltonian Baker's map. 
We can find a slightly different dimension which is sensitive to
the number of windings of the orbit around the space.
Numerically 
we generate the fixed points order by order instead so the area is
initially empty and fills with fixed points at each iteration.
We must get the same box-counting dimension for this set, that is, 
the set fills the allowed area, 
but the spectrum of weighted dimensions is different as shown below.

To obtain the dimensions numerically we isolate
the periodic orbits
order by order analytically.  The periodic orbits correspond to fixed
points of the map.  The 
points are then numerically 
covered with
boxes in the $(\theta,L)$ plane to explicitly determine the dimension.
Let $y=\exp{(i\theta_{n+1})}$, $x=\exp{(i\theta_n)}$, and 
$b_j=\exp{(i\pi j/4)}$.
With $a=\tanh(\tau_1/2)=\sqrt{2}(1+\sqrt{2})^{-1/2}$ 
the $\theta$ map eqn. (\ref{mapeq}) can be written as
	\be
	y={x+ab_j^{-1}\over 1+ab_jx} 
	\ \ 
	\ee	
for all points $(\theta,L)$ in the $j$th diagonal strip exiting the 
$j$th edge.
All $n=1$ fixed points satisfy $b_j^2y^2=1$ which implies
$\theta_*=-{j\pi \over 4}$ and $L=0$.
These correspond to the $4$ short period orbits and their inverses
shown in fig.\ \ref{fund}.

The fixed points at any order will satisfy an equation of the form
$y={(Ay+B)/( Cy+D)}$
and the periodic orbits are solutions to a quadratic equation.
The fixed points are found by using
the following recurrence relations:
	\ba
	A^{(n)} &=& A^{(n-1)}+ab_{j^{(n)}}^{-1} C^{(n-1)}\nonumber\\
	B^{(n)} &=& B^{(n-1)}+ab_{j^{(n)}}^{-1}D^{(n-1)}\nonumber\\
	C^{(n)} &=&C^{(n-1)}+ab_{j^{(n)}}A^{(n-1)}\nonumber\\
	D^{(n)} &=&D^{(n-1)}+ab_{j^{(n)}}B^{(n-1)} \  \ .
	\ea
At order $n=5$, there are  $\sum_{n=1}^{n=5} N(n)=19,624$ periodic points 
exactly as predicted by 
eqn.\ ({\ref{recur}).  These are
shown in fig.\ \ref{fractal}.

\begin{figure}[tbp]
\centerline{{
\psfig{file=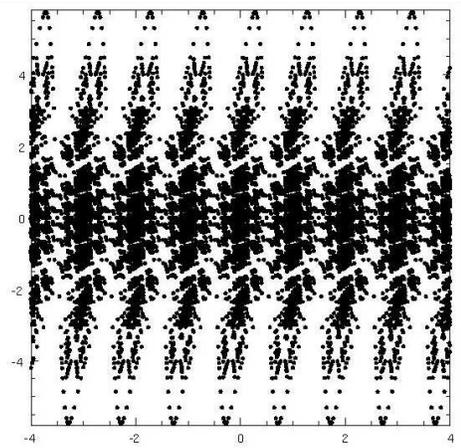,width=2.5in}}} \vskip 15truept
\caption{Each point corresponds to a periodic orbit.
All fixed points up to order $n=5$ are shown. There are $19,624$ in all.
}
\label{fractal}
\end{figure}

Numerically we count the number of boxes
needed to cover the set as $\epsilon\rightarrow 0$.
We measure the box-counting dimension as $2$, as predicted.
To find the weighted dimensions we let
$p_i$ equal the number of fixed points in a box and it is important to
note that 
we count each time a given fixed point occurs, so short orbits
are more heavily weighted.  We find the
information dimension is $D_1=1.4 \pm 0.1$.
The weight assigned to each box is sensitive to the number of windings
around the space and therefore measures a different topological
property than eqn. (\ref{dq}).

In the transition to
quantum mechanics, 
remnants of chaos
have been found in the form of 
smooth
scars of enhanced amplitude along the classical periodic orbits
\cite{heller1}.
We argue that the scars are the
residue of the most frequently encountered orbits in the fractal set.  

The semiclassical wave
function can be constructed as a superposition of modes 
$	\Psi(\vec x)=\sum_k \hat \phi_k  \psi_k(\vec x)$ 
with $\psi_k$ eigenvectors of the Laplace-Beltrami operator,
$\nabla^2 \psi_k=-k^2 \psi_k$.
The intuition that the quantum system blurs any chaotic features
derives from two important conjectures.  
The first conjecture asserts that the eigenvalues 
are given by
random matrix theory for systems which are classically completely
chaotic.  As a consequence, the amplitudes 
$\hat \phi_{\vec k}$ 
are drawn from a Gaussian
random ensemble with a flat spectrum and should be
thoroughly mixed and
featureless.  
The second conjecture, due to Berry \cite{berry}, 
implies that the eigenmodes will
be concentrated in the regions of phase space which are 
visited by the classical trajectories
given an infinite time.  A completely chaotic region of phase space is
covered by mixing trajectories and at first glance Berry's
conjecture seems to imply a uniform distribution.

While these arguments for uniformity are convincing, one might wonder
what happened to the structure of the periodic orbits.
It is no surprise that the periodic orbits, which play such a prominent
role in the classical dynamics, should also play a defining role in the
quantum system.  
Periodic loops have in fact been
the backbone of a quantization prescription in quantum chaos
\cite{{gutz1},{bt}}.   
Scars are the first evidence of quantum features related to these
closed orbits.
Heller discovered scars 
in the evolution of a wave-packet moving
along a short orbit in a chaotic stadium 
\cite{heller1}.
He deduced that only certain eigenstates would contribute to the wave
function as it traversed the orbit.
Since the wave-packet was peaked about a single periodic orbit, the
few contributing eigenstates must also show enhanced
intensity along that orbit.  The fewer the number of
eigenstates which contribute to the wave-packet, 
the larger the intensity of the scar along that loop.

Rather than being inconsistent with Berry's conjecture,
the scar can be interpreted as a consequence of it.
Scars occur in the regions of space where classical trajectories spend
the most time.
A typical trajectory will near a periodic orbit  and
then, due to the instability, will veer away onto a different periodic
orbit.  In this way, typical paths can be built from
segments of periodic orbits.
The most frequently visited periodic
orbits tend to scar the most substantially.
For a fully chaotic system, the fractal set of closed orbits will fill a
volume in phase space.
This would lead to
some uniformity in the distribution of classical orbits and hence
their quantum counterparts, as Berry suggested.
However, not all closed orbits are created equal.
Some may be visited by typical aperiodic trajectories more frequently
than others.

The scarring 
phenomenon has been investigated primarily in terms of the
dynamics of a wave-packet for test particle motion.  
Since the periodic orbits are unstable, a particle
will depart from that trajectory by a factor of
$\exp{(-\lambda t)}$ where $\lambda $ is the positive Lyapunov exponent.
The frequency $\omega $ with which the wave-packet traces out the path
must be sufficiently short, $\lambda/\omega \le 1$, if the orbit is to 
be susceptible to scarring.  For paths of length $\ell$, the frequency
is $\omega=2\pi/\ell$.  A hyperbolic surface is specified by $\lambda = 1$ in
curvature units
and the scarring condition is simply
$\ell\le 2\pi$ \cite{{heller1},{as}}.
Only the shortest orbits scar.  
The short orbits are also the ones visited most frequently. 
The paths shown in fig.\ \ref{fund} are all
of length $\ell\sim 3.06$ and are all potential avenues for scars.

On the compact octagon, 
the eigenmodes must obey precisely the same conditions as the
periodic orbits:
$\psi_k(\vec x)=\psi_k(g_j \vec x)$.  
The eigenmodes must be built out of
primitives which correspond to the primitive periodic geodesics.
For compact hyperbolic spaces, a relation between eigenmodes and
closed loops is evident.
Aurich and Steiner investigated 
scars in an asymmetric octagon in Refs
\cite{as}.
They confirmed 
that the weighted spectral density
of states shows bands related to the periodic recurrence of the wave
packet along the orbit.  The eigenstates under the
bands had de Broglie
wavelength,
$\lambda_B\sim 2\pi/k$, a multiple of the length of the orbit,
leading to the band quantization condition
$k\sim {2\pi n \over \ell}$
for integer band quantum number $n$ \cite{as}.
The fewer the number of states contributing to a band, the more likely
they were to show scars.
Consistent with this observation,
the authors found that very low
energy eigenstates did scar short orbits.  
The low energy eigenstates had few orbits to choose from and therefore
tended to show scars.
(They found scars can appear as underdensities as well as 
overdensities \cite{as}.)
For higher energy eigenstates however, they found
that scars were absent.  
High-$k$ states have many orbits to choose from and so are less
likely to scar along a given one.
In fact, since $D_q=2$, we know
that as $k\rightarrow \infty$, the area is covered by closed
orbits.  
So, even if a given high-$k$
eigenstate aligns with some closed orbits, it will be
entirely consistent with a random statistical distribution.

We should emphasize that a full description of the eigenmodes requires
an infinite sum over orbits, or at least a very large and cautiously
truncated sum.  Still, we can guess the location of scars in physical
space without this detailed description by exploiting the fractal connection.
The scarred features can be described qualitatively for any
classically chaotic system without having
to quantize the system: simply isolate
the most frequently visited orbits in the
fractal
set of periodic orbits.  These will be the site for quantum scars.
This offers a simple indicator of the quantum remnants of chaos.

A more strongly
multifractal set would show more extreme scars than the weak scars on
the octagon.  But the compact space 
is interesting as a model of the universe and, like our universe, is
suggestively marginal between homogeneity and inhomogeneity.
If the octagon were a $2d$
universe, created in a Big Bang, then
fluctuations in the geometry could be expanded in terms of the eigenmodes with
amplitudes drawn from a
random Gaussian ensemble with a flat spectrum.
The superposition of fluctuations would show
weak scars along short 
orbits from the low-$k$ modes but would appear increasingly random on
smaller and smaller scales.

Since galaxies have their origin in quantum fluctuations,
the distribution of large-scale structure would reflect
the homogeneity on nearly all scales except the largest where
the pattern of galaxies could carry memory of the quantum scars.
On scales comparable to the
area of the universe,
the distribution of matter could appear filamentary and web-like.

Our own $3d$ universe does show vast voids and huge filaments in the
distribution of large-scale structure.  
Galaxies would trace the scars
and chart
the 
shortest loops around a finite universe.
If our universe is finite and
negatively curved, the cobweb of luminous matter might be a residue
of primordial quantum scars.

\vskip 5pt

JDB and JL gratefully acknowledge PPARC support.
We thank R. Aurich for useful comments.

\end{document}